\documentclass[%
 reprint,
 amsmath,amssymb,
 aps,
]{revtex4-1}

\usepackage{graphicx}
\usepackage{dcolumn}% Align table columns on decimal point
\usepackage{bm}% bold math
\newcommand{\average}[1]{\ensuremath{\langle#1\rangle} }
\def\vector#1{\mbox{\boldmath $#1$}}

\begin{document}

\preprint{APS/123-QED}

\title{Spin-orbit-coupled ferroelectric superconductivity}% Force line breaks with \\
%% \thanks{A footnote to the article title}%

\author{Shota Kanasugi}
\thanks{kanasugi.shouta.62w@st.kyoto-u.ac.jp}
\author{Youichi Yanase}%
\affiliation{%
 Depertment of Physics, Graduate School of Science, Kyoto University, Kyoto 606-8502, Japan
}%
\date{\today}% It is always \today, today,

\begin{abstract}
Motivated by recent studies on ferroelectric-like order coexisting with metallicity, we investigate ferroelectric (FE) superconductivity in which a FE-like structural phase transition occurs in the superconducting state. 
We consider a two-dimensional $s$-wave superconductor with Rashba-type antisymmetric spin-orbit coupling (ASOC). 
Assuming linear relationship between polar lattice displacement and strength of the ASOC, we treat the Rashba-type ASOC as a molecular field of FE-like order. 
It is shown that the FE-like order is induced by the magnetic field when the system is superconducting. 
Furthermore, we clarify the FE superconductivity in a low carrier density regime, which was recently discovered in doped SrTiO$_3$. 
It is demonstrated that the FE superconducting state can be stable in this regime in the absence of the magnetic field. 
Our results open a way to control the electric polarization by superconductivity, that is, superconducting multiferroics. 
\end{abstract}

\pacs{Valid PACS appear here}% PACS, the Physics and Astronomy
\maketitle

%\tableofcontents

%%%% Sec. I
\section{\label{sec:level1}Introduction}
% Ferrolectric metal
Recent experimental observations have generated a flurry of interest in the relationship between metallicity, ferroelectricity and superconductivity \cite{NatMater.12.1024-1027, NatPhys.13.643-648, PhysRevLett.119.207001}. 
In principle, metals cannot exhibit ferroelectricity because dielectric polarization is screened by conduction electrons. 
However, in 1965, Anderson and Blount predicted the existence of ferroelectric (FE) metals in which FE-like structural phase transition occurs in the metallic state \cite{PhysRevLett.14.217}. 
A lot of experiments have been devoted to searches of FE metal for half a century \cite{RevModPhys.47.637, paduani2008martensitic, PhysRevLett.92.065501, PhysRevB.82.054108, PhysRevLett.104.147602, PhysRevB.84.064125, kim2016polar}, and recently, a FE-like structural phase transition was observed in metallic LiOsO$_3$ \cite{NatMater.12.1024-1027}. 

% Ferroelectric superconductor
Following the discovery of a candidate of FE metal, the relationship between ferroelectricity and superconductivity has also received a lot of attention. 
A promising candidate material of FE superconductivity, in which FE-like order coexists with superconductivity, is SrTiO$_3$ (STO). 
STO is a quantum paraelectric (PE) \cite{PhysRevB.19.3593} whose dielectric constant is extremely high about 20,000 at low temperatures, but a pure compound does not exhibit ferroelectricity because the long-range FE order is suppressed by quantum fluctuation. 
However, STO turns into a  FE by tiny isovalent Ca doping \cite{PhysRevLett.52.2289} , isotopic substitution of $^{16}$O with $^{18}$O \cite{PhysRevLett.82.3540} or application of stress \cite{PhysRevB.13.271}. 
On the other hand, charge carriers can be doped into the STO by substituting Sr atoms with La, Ti with Nb or by oxygen vacancy, and then, STO becomes a metal showing superconducting instability at low temperatures \cite{PhysRevLett.12.474}. 
Based on these observations, Rischau {\it et al}. \cite{NatPhys.13.643-648} have recently performed various measurements for Sr$_{1-x}$Ca$_x$TiO$_{3-\delta}$ with oxygen vacancy and weak Ca-doping. 
They mapped out a temperature-versus-carrier density phase diagram in which a coexistent phase of FE-like order and superconductivity exists. 
Their experimental results suggest the FE superconductors. 

% Non-centrosymmetric superconductor
FE-like structural phase transition in the metallic state can be defined as the appearance of polar axis and the disappearance of inversion center in the presence of conducting electrons \cite{NatMater.12.1024-1027}. 
In such a situation, antisymmetric spin-orbit coupling (ASOC) is induced by the combination of spin-orbit coupling and spontaneous inversion symmetry breaking. 
This means that the emergence of the FE-like order in a spin-orbit coupled system is characterized by the appearance of ASOC \cite{PhysRevLett.115.026401}. 
Recent theoretical and experimental studies have elucidated various intriguing phenomena of noncentrosymmetric (NCS) superconductors, for example, parity mixing of Cooper pairs \cite{NCS-SC.Springer}, upper critical field exceeding the Pauli-Chandrasekhar-Clogston limit \cite{fujimoto2007electron, PhysRevLett.92.097001, saito2016superconductivity} and stabilization of helical superconducting state under a magnetic filed \cite{Dimitrova2003, PhysRevB.70.104521, PhysRevB.75.064511, PhysRevLett.111.057005}. 
FE superconductors may be a new platform of such exotic superconducting states. 
However, relations between the spontaneous FE-like order and NCS superconductivity remain a mystery. 

% Dilute superconductivity
The pairing mechanism of superconductivity in STO is also mysterious despite a long history of investigations. 
Metallic STO exhibits superconductivity at very low carrier density of the order of 10$^{17}$ cm$^{-3}$ \cite{PhysRevX.3.021002, PhysRevLett.112.207002}, which is exceptionally low compared to any other known superconductors. 
In this region of carrier density, the Fermi temperature $T_{\rm F}$ is lower than the Debye temperature $T_{\rm D}$ \cite{BURNS1980811}, and it seems difficult to generate Cooper pairs by phonon-mediated attractive interactions \cite{RevModPhys.62.1027, PhysRevB.94.224515}. 
Recent theories propose several situations to explain the pairing mechanism of this extremely dilute superconductivity in STO. 
For example, Edge {\it et al}. \cite{PhysRevLett.115.247002} have proposed dilute superconductivity induced by the quantum FE fluctuations. 
Despite various theoretical works \cite{PhysRevLett.115.247002, PhysRevB.94.224515, takada1978plasmon, takada1980theory, Gorkov4646, arce2018quantum, arxiv08121, kedem2018analytical, wolfle2018superconductivity}, there is still no consensus about the origin of superconductivity in STO. 
However, recent works point to the superconductivity influenced by the FE quantum criticality. 

% Our research
In this paper, we investigate a ubiquitous mechanism of the FE superconductivity in a spin-orbit coupled system, namely, {\it spin-orbit-coupled FE superconductivity}. 
We consider a two-dimensional (2D) $s$-wave Rashba superconductivity on a square lattice as a minimal model, and analyze the model with the use of the mean-field (MF) theory. 
Assuming linear relationship between polar lattice distortion and the induced Rashba-type ASOC, we treat coupling strength of the ASOC as a FE order parameter. 
Then, the free energy including the energy of polar lattice displacement is minimized with respect to the superconducting and FE order parameters. 
We map out the phase diagram for the thermodynamic states. 
The results reveal two carrier  density regions distinguished by the stability of FE superconductivity, {\it i.e.} high carrier density regime and low carrier density regime. 
In the high carrier density regime near a FE critical point, it is shown that the FE superconducting state is stabilized under the applied magnetic field, even though it is unstable at zero magnetic field. 
On the other hand, in the low carrier density regime, the FE superconducting state can be stabilized without applying the magnetic field, owing to the enhancement of density of states (DOS) by the FE-like order. 
The result is consistent with the experimental phase diagram of Sr$_{1-x}$Ca$_x$TiO$_{3-\delta}$ \cite{NatPhys.13.643-648}, in which the FE superconducting state is stabilized in the low carrier density regime at zero field. 
Furthermore, it is implied that the FE superconducting phase may be stabilized in various superconductors by applying the magnetic field. 
Such effect arises from an unusual magnetic response of NCS superconductors \cite{NCS-SC.Springer, fujimoto2007electron, PhysRevLett.92.097001, saito2016superconductivity}. 
Thus, not only the FE-like structural transition, but also other types of spontaneous inversion symmetry breaking \cite{PhysRevLett.115.207002} might be induced in spin-orbit coupled superconductors in a controllable way. 
Previous works have not focused on the feedback effect of superconductivity on the inversion-symmetry-breaking order. 
Thus, our theoretical proposal might broaden the research field of the superconductivity, ferroelectricity, and higher order multipole order. 

%%%%%%%%%%%%%%%% Figure of Model %%%%%%%%%%%%%%%%%%%%%%%%%%%%%%%
\begin{figure}[b]
\centering
    \includegraphics[width=90mm,clip]{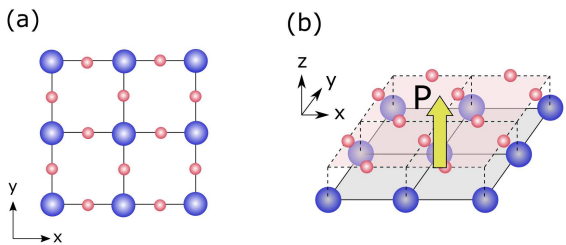}
  \caption{\label{fig:polar} 
(a) Crystal structure of a 2D tetragonal lattice. 
Blue and red circles show the metal ions and the non-metal ions, respectively. 
(b) Polar lattice displacement of the 2D tetragonal system. 
 Displacement of the metal (light blue plane) and non-metal (light red plane) planes induces the FE-like order, and then, the Rashba-type ASOC arises. 
}
\end{figure}
%%%%%%%%%%%%%%%%%%%%%%%%%%%%%%%%%%%%%%%%%%%%%%%%%%%%%%%%%%%%

% Construction
This paper is constructed as follows. 
In Sec. \ref{sec:level2}, a model of electron-lattice coupled 2D $s$-wave Rashba superconductor is introduced. 
We analyze the model with the use of MF theory, and then, the energy of polar lattice distortion is included phenomenologically. 
In Sec. \ref{sec:level3}, we show the results of high carrier density regime. 
It is shown that the FE superconducting state is stabilized in the presence of the magnetic field, although it is unstable at zero field. 
In Sec. \ref{sec:level4}, a significantly different relationship between ferroelectricity and superconductivity in the low carrier density region is demonstrated. 
We show that dilute superconductors such as STO are advantageous to stabilize the FE superconducting state, and thus, the coexistent FE superconducting state is stabilized even at zero magnetic field. 
A possibility of a self-organized topological superconducting state by FE-like structural transition is also discussed. 
In Sec. \ref{sec:level5}, an qualitative understanding of our results is given by the Ginzburg-Landau free energy. 
Finally, a brief summary and discussion are given in Sec. \ref{sec:level6}.

%%%% Sec. II
\section{\label{sec:level2}Model and Formulation}
%% Model
\subsection{\label{sec:level2A}Model} 
In this work we consider a minimal model for a spin-orbit coupled FE superconductivity, that is, a 2D electron system on a square lattice coupled to a polar lattice distortion [Fig. \ref{fig:polar}(a)]. 
This kind of 2D superconductivity is actually realized in the $\delta$-doped STO \cite{Nature.462.487}, for example. 
In this model, a FE-like structural phase transition with spontaneous mirror symmetry breaking [Fig. \ref{fig:polar}(b)] is taken into account. 
The crystallographic point group $D_{4h}$ descends to the subgroup $C_{4v}$ by the FE-like structural transition. 
Hence, due to a spin-orbit coupling, the Rasba-type ASOC is induced by the FE-like structural transition \cite{Rashba}. 
Based on these observations, we introduce a model of electron-lattice coupled 2D Rashba superconductor as follows:
\begin{eqnarray}
\mathcal{H}&=&\mathcal{H}_{\rm kin}+\mathcal{H}_{\rm pol}+\mathcal{H}_{\rm pair}+\mathcal{H}_{\rm Z} , \label{eq:1} \\
\mathcal{H}_{\rm kin}&=&\sum_{{\bm k}, s} \left[\varepsilon(\vector{k}) -\mu \right] c_{{\bm k}s}^{\dag}c_{{\bm k}s} , \\
\mathcal{H}_{\rm pol}&= & \sum_{{\bm k}, s, s'} \alpha\vector{g}(\vector{k})\cdot\hat{\vector{\sigma}}_{ss'}c_{{\bm k}s}^{\dag}c_{{\bm k}s'} , \\ 
 \mathcal{H}_{\rm pair} &=& \frac{1}{N}\sum_{{\bm k}, {\bm k}', {\bm q}} V(\vector{k}, \vector{k}')
c_{{\bm k}\uparrow}^{\dag}
c_{-{\bm k}+{\bm q}\downarrow}^{\dag}
c_{-{\bm k}'+{\bm q}\downarrow}
c_{{\bm k}'\uparrow} ,  \\
\mathcal{H}_{\rm Z} &=&-\mu_{\rm B}\sum_{{\bm k}, s, s'}\vector{H}\cdot\hat{\vector{\sigma}}_{ss'}c_{{\bm k}s}^{\dag}c_{{\bm k}s'} ,
\end{eqnarray}
where $c_{{\bm k}s}$ is the annihilation operator of an electron with momentum $\vector{k}$ and spin $s=\uparrow, \downarrow$. 
This model represents the Hamiltonian of electrons, and later we add the energy arising form the polar lattice distortion. 

The first term $\mathcal{H}_{\rm kin}$ is the kinetic energy term including the chemical potential $\mu$. 
We assume a simple nearest-neighbor hopping tight-binding form $\varepsilon(\vector{k})=-2t(\cos k_x+\cos k_y)$ on a square lattice, 
where $t$ is the hopping integral. 
We choose the unit of energy as $t=1$. 

The second term $\mathcal{H}_{\rm pol}$ represents the Rashba-type ASOC which is induced by the FE-like structural phase transition. 
The $g$-vector of Rashba-type ASOC on a square lattice is described as $\vector{g}(\vector{k})=(-\sin{k_y}, \sin{k_x}, 0)$. 
Here, we assume $P \propto \alpha$, with $P$ being a lattice polarization [Fig. \ref{fig:polar}(b)] and $\alpha$ is the coupling strength of ASOC. 
This relation has been verified by a first-principles calculation of perovskite 2D electron systems \cite{PhysRevB.88.041302}, which shows that the Rashba-type ASOC is induced by the polar lattice displacement. 
Based on this picture, we treat $\alpha$ as an order parameter which characterizes the FE-like order.  

The third term $ \mathcal{H}_{\rm pair}$ expresses the attractive interaction for superconductivity. 
$N=L^2$ is the number of sites, and $\vector{q}$ is the center-of-mass momentum of Cooper pairs. 
We consider the $s$-wave superconductivity expected in bulk STO \cite{PhysRevLett.45.1352, PhysRevB.92.174504} and other conventional superconductors by adopting momentum-independent pairing interaction $V(\vector{k}, \vector{k}')=-V_s$. 
For NCS superconductors, the parity mixing of the superconducting gap function generally occurs \cite{NCS-SC.Springer}. 
Thus, in this case, an odd-parity $p$-wave component is induced in addition to the $s$-wave component \cite{NCS-SC.Springer, RevModPhys.63.239}. 
However, when the pairing interaction in the $s$-wave channel is dominant, FE-like order is hardly affected by the $p$-wave component. 
Therefore, we neglect the $p$-wave gap function in the following. 

In this work we clarify the magnetic response of FE superconductor. 
For this purpose, we introduce the Zeeman coupling term $\mathcal{H}_{\rm Z}$ by the fourth term of 
Eq. (\ref{eq:1}), where $\hat{\vector{\sigma}}$ is the Pauli matrix and $\mu_{\rm B}$ is the Bohr magneton. 
We consider two directions of the magnetic field, 
{\it i.e.} the perpendicular field $\vector{H}=(0, 0, H_z)$ and the in-plane field $\vector{H}=(H_x, 0, 0)$. 
In most bulk superconductors, the orbital depairing effect is not negligible, and therefore, we should include the gauge interaction with vector potential in addition to the Zeeman coupling term. 
However, the orbital depairing effect is suppressed by the geometry when we consider quasi-2D system in the in-plane magnetic field, and we can ignore it. 
For the perpendicular magnetic field, the importance of the orbital depairing effect depends on the Maki parameter $\alpha_{\rm M}=\sqrt{2}H_{\rm c2}^{\rm orb}/H_{\rm c2}^{\rm P}$, where $H_{\rm c2}^{\rm orb}$ is the orbital limiting field and $H_{\rm c2}^{\rm P}$ is the Pauli limiting filed. 
Since $\alpha_{\rm M} \propto |\Delta|/E_{\rm F}^{*}$ with $E_{\rm F}^{*}$ being a renormalized Fermi energy, heavy fermion superconductors or low carrier density superconductors with a small $E_{\rm F}^{*}$ are candidates of our analysis, for instance. 
More precise study including the orbital depairing effect is left for a future work. 

In the Rashba superconductors, a Fulde-Ferrell-Larkin-Ovchinnikov (FFLO) state \cite{PhysRev.135.A550,  LO} is stable and the center-of-mass momentum $\vector{q}$ is finite when the in-plane filed is applied \cite{Dimitrova2003, PhysRevB.70.104521, PhysRevB.75.064511}. 
However, because the phase diagram of FE superconductivity is hardly affected by a finite $\vector{q}$, we fix $\vector{q}=\vector{0}$, {\it i.e.} the Bardeen-Cooper-Schrieffer (BCS) state in the following. 
A possibility of the FE FFLO state is briefly discussed later. 
It is expected that some ubiquitous relationships between ferroelectricity and superconductivity are captured by our model. 
Although the bulk STO is a three-dimensional and multi-orbital \cite{nakamura2013multi} system, a material-specific study taking into account these ingredients is left for a future work. 

%% Mean-field theory
\subsection{\label{sec:level2B}Mean-field theory}
We investigate the superconducting state by means of mean-field   theory.
The pairing interaction term $\mathcal{H}_{\rm pair}$ is approximated as follows:
\begin{eqnarray}
\mathcal{H}_{\rm pair} &=& 
 -\frac{V_s}{N}\sum_{{\bm k}, {\bm k}'}
c_{{\bm k}\uparrow}^{\dag}
c_{-{\bm k}\downarrow}^{\dag}
c_{-{\bm k}'\downarrow}
c_{{\bm k}'\uparrow} \nonumber \nonumber\\
&\simeq& \sum_{{\bm k}}\left( \Delta c_{{\bm k}\uparrow}^{\dag}c_{-{\bm k}\downarrow}^{\dag}+{\rm h.c.} \right)
+\frac{N}{V_s}|\Delta|^2 ,
\end{eqnarray}
by introducing the order parameter
\begin{equation}
\Delta= -\frac{V_s}{N}\sum_{{\bm k}'}\average{c_{-{\bm k}'\downarrow}c_{{\bm k}'\uparrow}} \label{eq:gap1} .
\end{equation}

To describe the MF Hamiltonian in a matrix form, we here define the vector operator
\begin{equation}
\hat{C}^{\dag}_{\bm k}=
\left(
c_{{\bm k}\uparrow}^{\dag}, 
c_{{\bm k}\downarrow}^{\dag}, 
c_{-{\bm k}\uparrow}, 
c_{-{\bm k}\downarrow}
\right) .
 \end{equation}
Then, we obtain 
\begin{equation}
\mathcal{H}_{\rm MF}= \frac{1}{2}\sum_{{\bm k}}\hat{C}^{\dag}_{\bm k} \hat{\mathcal{H}}_{\rm BdG}(\vector{k}) \hat{C}_{\bm k} + E_{\rm c} ,
\end{equation}
where $E_{\rm c}$ is 
\begin{equation}
E_{\rm c} =\sum_{{\bm k}}[\varepsilon(\vector{k})-\mu]+\frac{N}{V_s}|\Delta|^2 .
\end{equation}
The Bogoliubov-de Gennes (BdG) Hamiltonian $\hat{\mathcal{H}}_{\rm BdG}(\vector{k})$ is given by
\begin{equation}
 \hat{\mathcal{H}}_{\rm BdG}(\vector{k})=
    \begin{pmatrix}
       \hat{\mathcal{H}}_0(\vector{k}) & \hat{\Delta}  \\ 
     \hat{\Delta}^{\dag} &  -\hat{\mathcal{H}}_0^{\rm T}(-\vector{k})
    \end{pmatrix} , 
\end{equation}
where
\begin{equation}
\hat{\mathcal{H}}_0(\vector{k}) = [\varepsilon(\vector{k})-\mu ] \hat{\sigma}_0+[\alpha\vector{g}(\vector{k})-\mu_{\rm B} \vector{H}]\cdot\hat{\vector{\sigma}}
\end{equation}
and $\hat{\Delta} = \Delta i \hat{\sigma}_{y}$. 

We carry out Bogoliubov transformation by using the unitary matrix 
$\hat{U}(\vector{k})$:
\begin{eqnarray}
\mathcal{H}_{\rm MF} &=& \frac{1}{2}\sum_{{\bm k}}\hat{C}^{\dag}_{\bm k} \hat{U}(\vector{k}) \hat{U}^{\dag}(\vector{k}) 
\hat{\mathcal{H}}_{\rm BdG}(\vector{k}) \hat{U}(\vector{k})\hat{U}^{\dag} (\vector{k}) \hat{C}_{\bm k} +E_{\rm c} \nonumber \\
&=& \frac{1}{2}\sum_{{\bm k}}\hat{\Gamma}^{\dag}_{\bm k} \hat{E}_{\rm BdG}(\vector{k}) \hat{\Gamma}_{\bm k} + E_{\rm c}. 
\end{eqnarray}
The Bogoliubov quasiparticle operator 
$\hat{\Gamma}^{\dag}_{\bm k} = \hat{C}^{\dag}_{\bm k} \hat{U}(\vector{k})$ 
and the diagonal matrix $\hat{E}_{\rm BdG}(\vector{k})$ are expressed with using pseudospin $\tau = \uparrow, \downarrow$:
\begin{eqnarray}
\hat{\Gamma}^{\dag}_{\bm k} &=& 
\left(
 \gamma_{{\bm k}\uparrow}^{\dag}, \gamma_{{\bm k}\downarrow}^{\dag}, 
 \gamma_{-{\bm k}\uparrow}, \gamma_{-{\bm k}\downarrow}
\right) ,  \\
\hat{E}_{\rm BdG}(\vector{k}) &=& {\rm diag}
\left( E_{{\bm k} \uparrow}, E_{{\bm k} \downarrow},
 -E_{-{\bm k} \uparrow}, -E_{-{\bm k} \downarrow} \right) .
\end{eqnarray}
Then, the MF Hamiltonian is described as 
\begin{equation}
\mathcal{H}_{\rm MF}= \sum_{{\bm k},\tau} E_{{\bm k}\tau}
\left( \gamma_{{\bm k}\tau}^{\dag} \gamma_{{\bm k}\tau} -\frac{1}{2} \right) + E_{\rm c} .
\label{eq:MF_Bogo}
\end{equation}
Because of the particle-hole symmetry of the BdG Hamiltonian, the unitary matrix $\hat{U}(\vector{k})$ is expressed as follows:
\begin{equation}
\hat{U}(\vector{k}) = 
\begin{pmatrix}
u_{{\bm k}\uparrow}^{(\uparrow)} & u_{{\bm k}\uparrow}^{(\downarrow)}
& -v_{-{\bm k}\uparrow}^{(\uparrow)*} & -v_{-{\bm k}\uparrow}^{(\downarrow)*} \\[2pt]
u_{{\bm k}\downarrow}^{(\uparrow)} & u_{{\bm k}\downarrow}^{(\downarrow)}
& -v_{-{\bm k}\downarrow}^{(\uparrow)*} & -v_{-{\bm k}\downarrow}^{(\downarrow)*} \\[2pt]
v_{{\bm k}\uparrow}^{(\uparrow)} & v_{{\bm k}\uparrow}^{(\downarrow)}
& u_{-{\bm k}\uparrow}^{(\uparrow)*} & u_{-{\bm k}\uparrow}^{(\downarrow)*} \\[2pt]
v_{{\bm k}\downarrow}^{(\uparrow)} & v_{{\bm k}\downarrow}^{(\downarrow)} 
& u_{-{\bm k}\downarrow}^{(\uparrow)*} & u_{-{\bm k}\downarrow}^{(\downarrow)*}
\end{pmatrix} 
\label{eq:unitary}.
\end{equation}
Thus, from Eqs. (\ref{eq:gap1}) and (\ref{eq:unitary}), the order parameter is obtained by
\begin{equation}
\Delta = - \frac{V_s}{N} \sum_{{\bm k},s,\tau} 
s v_{{\bm k}\overline{s}}^{(\tau)*} u_{{\bm k}s}^{(\tau)} f(s E_{{\bm k}\tau}) ,
\label{eq:gap2}
\end{equation}
where $f(E)$ is the Fermi-Dirac distribution function.
Equation (\ref{eq:gap2}) is the gap equation to be solved numerically. 
Since we treat the Rashba coupling constant $\alpha$ as an order parameter which describes the FE-like order, and $\alpha$ causes a change in the band structure, we have to solve the particle number equation simultaneously. 
The particle number equation is obtained
\begin{equation}
n=\frac{1}{N}\sum_{{\bm k}, s} \average{c_{{\bm k}s}^{\dag} c_{{\bm k}s} } \label{eq:num1} ,
\end{equation}
where $n$ is the carrier density. 
In the basis of the Bogoliubov quasiparticle, Eq. (\ref{eq:num1}) is recast
\begin{equation}
n = \frac{1}{N} \sum_{{\bm k},s,\tau}
\left[
 |u_{{\bm k}s}^{(\tau)}|^2 f(E_{{\bm k}\tau}) 
+ |v_{{\bm k}s}^{(\tau)}|^2 f(-E_{{\bm k}\tau})
\right] .
 \label{eq:num2}
\end{equation}
From Eq. (\ref{eq:MF_Bogo}), the Helmholtz free energy per site $f_{\rm el}$ is obtained as
\begin{equation}
f_{\rm el} = -\frac{1}{N\beta}\sum_{{\bm k}, \tau}
\left[ \ln \left( 1 + e^{-\beta E_{{\bm k}\tau}} \right) + \frac{\beta E_{{\bm k}\tau}}{2} \right]
+ \frac{E_{\rm c}}{N} +\mu n ,
\label{eq:free_BdG}
\end{equation}
where $\beta = 1/k_{\rm B} T$ is the inverse temperature.
Here, the last term of Eq. (\ref{eq:free_BdG}) is required when the carrier density of this system is fixed as $n$. 
Using $\Delta$ and $\mu$ obtained by solving Eqs. (\ref{eq:gap2}) and (\ref{eq:num2}), we calculate the electronic part of the free energy $f_{\rm el}$ from Eq. (\ref{eq:free_BdG}). 

%% Lattice polarization
\subsection{\label{sec:level2C}Lattice polarization}
To study the stability of FE superconducting state, we have to include the contribution of polar lattice displacement to the free energy. 
As explained in Sec. \ref{sec:level2A}, we assume a linear relation 
$P = C \alpha$, where $P$ and $C$ is the lattice polarization and the 
proportional constant, respectively. 
This allows us to introduce the energy loss by lattice polarization as follows:
\begin{equation}
f_{\rm pol} = \frac{1}{2}\gamma P^2 + \eta P^4 , 
\label{eq:free_pol}
\end{equation}
where $\gamma$ and $\eta$ are coefficients which describe the elasticity of the lattice. 

In globally NCS superconductors, it is known that the strength of ASOC satisfies $\alpha \ll E_{\rm F}$ \cite{NCS-SC.Springer} in most cases, where $E_{\rm F}$ is the Fermi energy. 
The second term of Eq. (\ref{eq:free_pol}) is included to cutoff the value of $\alpha$ in this realistic regime.

%%%%%%%%%%%%%%%% QFE %%%%%%%%%%%%%%%%%%%%%%%%%%%%%%%
\begin{figure}[b]
 \centering
 \includegraphics[width=95mm]{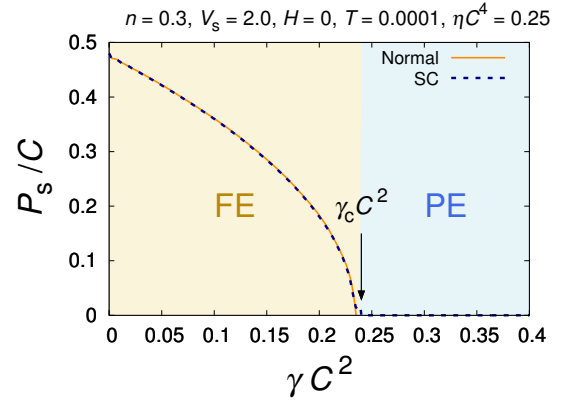}
 \caption{\label{gamma_metal} 
$\gamma$ dependence of the FE order parameter $P_s$ in the high carrier density regime ($n=0.3$). 
The temperature is set to $T=0.0001$, {\it i.e.} almost zero temperature. 
The blue dot line shows $P_s$ in the superconducting state, while the orange solid line shows $P_s$ in the normal state. 
The two lines almost coincide with each other. 
Yellow and light blue areas represent the FE regime ($P_s \neq 0$) and PE regime ($P_s = 0$), respectively. 
}
\end{figure}
%%%%%%%%%%%%%%%%%%%%%%%%%%%%%%%%%%%%%%%%%%%%%%%%%%%%%%%%%%%%

%% states
\subsection{\label{sec:level2D}Superconducting states}
The total free energy of the system is given by
\begin{equation}
f_{\rm tot} = f_{\rm el} + f_{\rm pol} . \label{eq:free_tot}
\end{equation}
The thermodynamically stable state is determined by minimizing the free energy $f_{\rm tot}$ with respect to the order parameter of superconductivity $\Delta$ and that of the ferroelectricity $P$. 
We here summarize the two superconducting states which can be stabilized in our model: the PE superconducting state, and the FE superconducting state. 
In the PE superconducting state, there is no polar lattice displacement  and 
$P_s = 0$, where $P_s$ is the lattice polarization minimizing the free energy. 
The FE superconducting state is the self-organized NCS superconducting state with 
$P_s \neq 0$. 
In the following discussion, we assume that the normal state is PE near the critical point of the FE order. 
This situation is realized in the carrier doped STO. 
Then, we elucidate the relationship between superconductivity and FE-like order. 

%%%%%%%%%%%%%%%% Phase diagram %%%%%%%%%%%%%%%%%%%%%%%%%%%%%%%
\begin{figure*}[t]
\begin{minipage}[t]{0.45\linewidth}
  \centering
 \includegraphics[width=100mm]{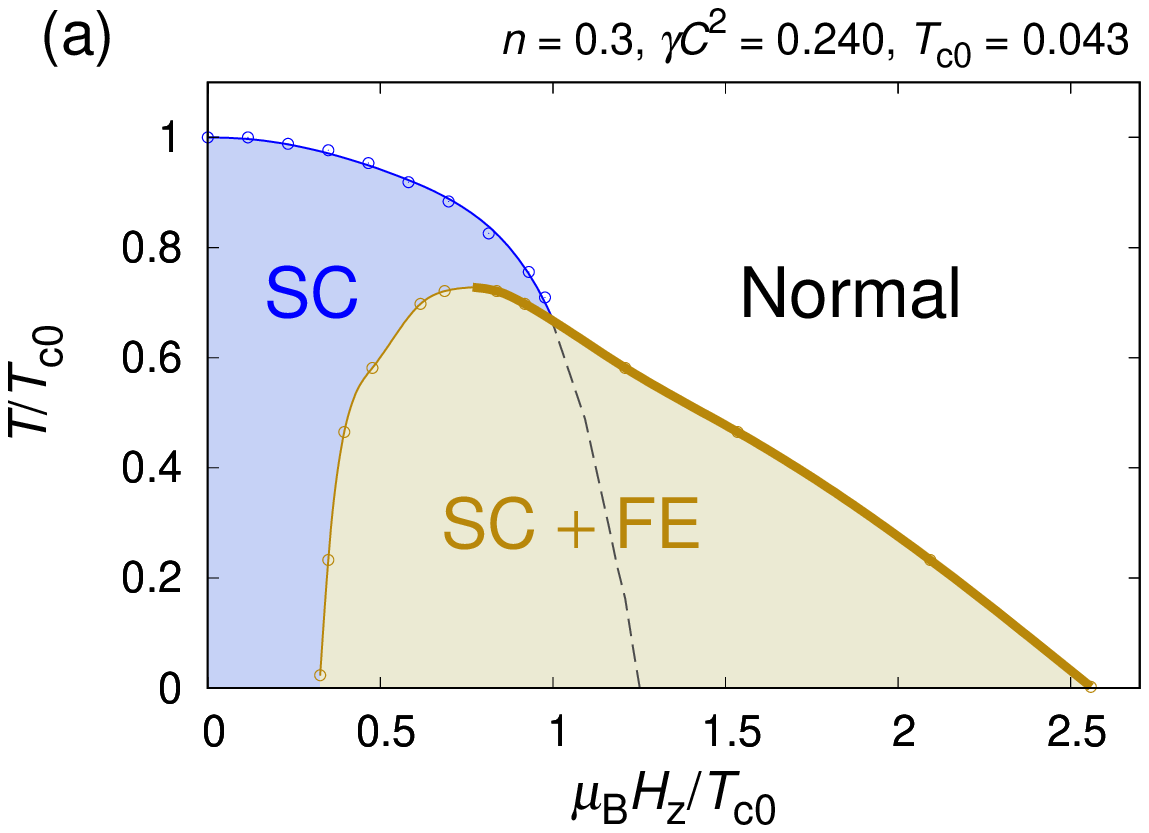}
\end{minipage}
\begin{minipage}[t]{0.45\linewidth}
\centering
 \includegraphics[width=100mm]{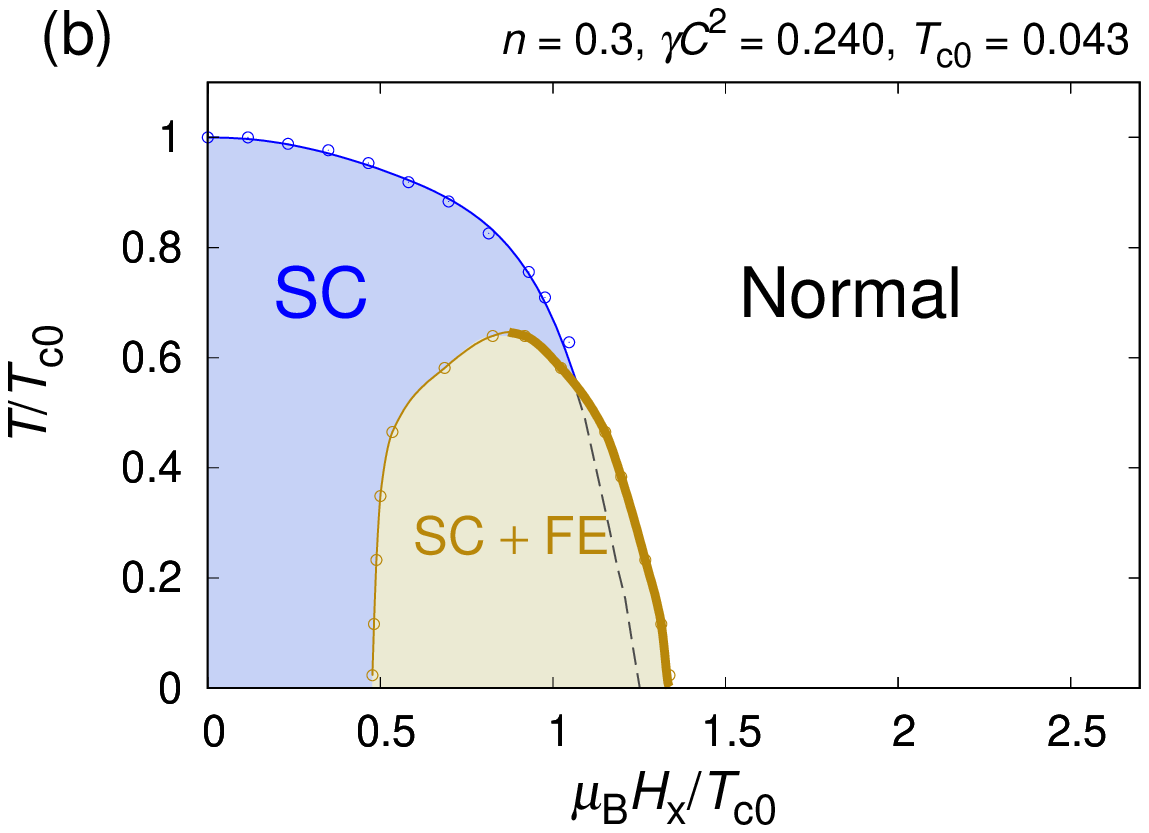}
\end{minipage}
 \caption{\label{phase_metal} 
 Superconducting phase diagram in the high carrier density regime for (a) perpendicular field, and (b) in-plane field. 
The bold (thin) line shows the first-order (second-order) phase transition line. 
The black dashed line shows a superconducting phase transition line of a 2D superconductor with inversion symmetry. 
The temperature $T$ and magnetic field $\mu_{\rm B}H$ are normalized by the superconducting transition temperature $T_{\rm c 0}$ at zero magnetic field. 
}
\end{figure*}
%%%%%%%%%%%%%%%%%%%%%%%%%%%%%%%%%%%%%%%%%%%%%%%%%%%%%%%%%%%%

%%%% Sec. III
\section{\label{sec:level3}High carrier density regime}
\subsection{\label{sec:level3A}Quantum ferroelectric criticality}

First, we examine the stability of FE superconducting state in the high carrier density regime by setting $n = 0.3$. 
The attractive interaction is chosen to be $V_s/t =2.0$, hence the superconducting transition temperature is  set to $T_{c0} = 0.043$ in the absence of the magnetic field. 
The following results are not qualitatively altered by other choice of $V_s$. 
To calculate the free energy, we have to determine the value of $\gamma$ and $\eta$. 
The cutoff vale $\eta$ is chosen to be $\eta = 0.25 /C^4$ so that 
$\alpha_s = P_s/C < 0.5$ is satisfied. 
A more important parameter, $\gamma$ is assumed so as to express the quantum PE state which realizes in STO for instance. 
Figure \ref{gamma_metal} plots the $\gamma$ dependence of $P_s$ at zero magnetic field and almost zero temperature. 
We see that $P_s$ vanishes at $\gamma=\gamma_{\rm c}=0.2396/ C^2$, and the quantum critical point (QCP) $\gamma_{\rm c}$ separates the FE phase and PE phase. 
To consider the quantum PE state in the vicinity of the QCP, we assume 
$\gamma=0.240 /C^2$, which is slightly larger than $\gamma_{\rm c}$, in the following.

\subsection{\label{sec:level3B}Perpendicular magnetic field}
\begin{figure}[t]
 \begin{center}
 \includegraphics[width=87mm]{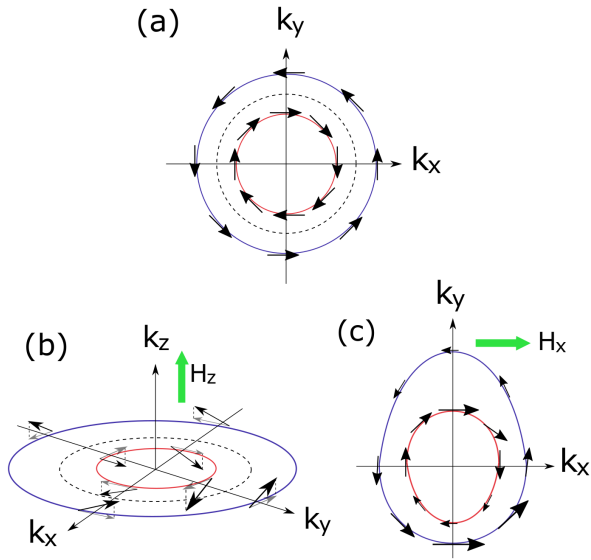}
 \end{center}
 \caption{\label{FS} 
 Split Fermi surfaces of a polar electron system with Rashba-type ASOC under 
(a) zero magnetic field, (b) perpendicular magnetic field, and (c) in-plane magnetic field. 
Arrows on the Fermi surfaces illustrate spin texture. 
}
\end{figure}

Figure \ref{phase_metal}(a) shows the $H_z$-$T$ phase diagram in the high carrier density regime. 
It is shown that the FE superconducting state is stabilized by the applied perpendicular magnetic field, although it is unstable at zero magnetic field. 
At the same time, the Pauli limiting field is drastically enhanced by the transition from the PE superconducting state to the FE superconducting state. 
These results can be understood on the basis of the spin susceptibility and spin texture of NCS superconductors \cite{fujimoto2007electron}. 
As shown in Figs. \ref{FS}(a) and (b), Cooper pairs formed by electrons with momentum $\vector{k}$ and $-\vector{k}$ ({\it i.e.} the BCS pairing) are possible under the perpendicular field for Rashba superconductors when 
$\mu_{\rm B}H_z \ll \alpha$ is satisfied. 
Thus, the Pauli depairing effect is suppressed due to the FE-like structural phase transition inside the superconducting state. 
Hence, the FE superconducting state is stable under the Zeeman magnetic field, and the Pauli limiting field is enhanced compared to the superconductors with inversion symmetry. 

\begin{figure}[ht]
  \begin{minipage}{0.75\linewidth}
    \centering
    \includegraphics[width=90mm,clip]{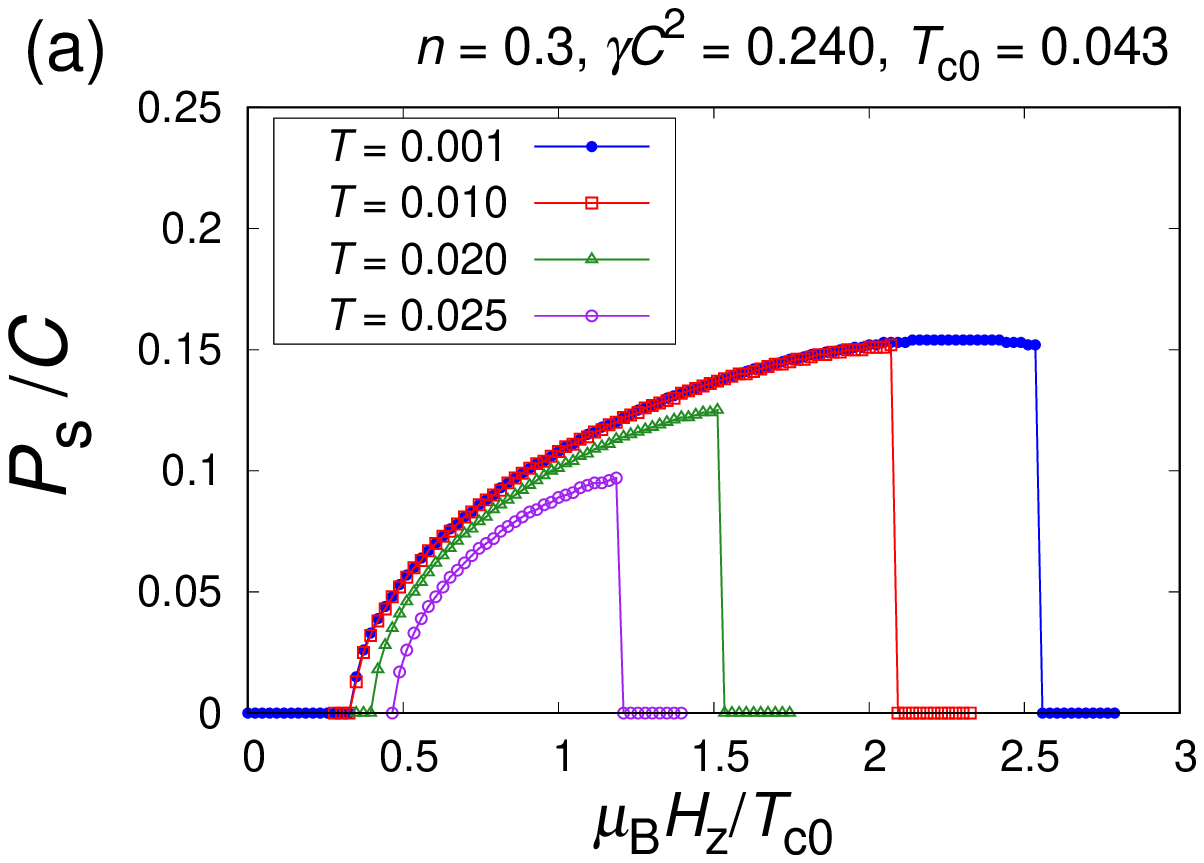}
  \end{minipage}
  \begin{minipage}{0.75\linewidth}
    \centering
    \includegraphics[width=90mm,clip]{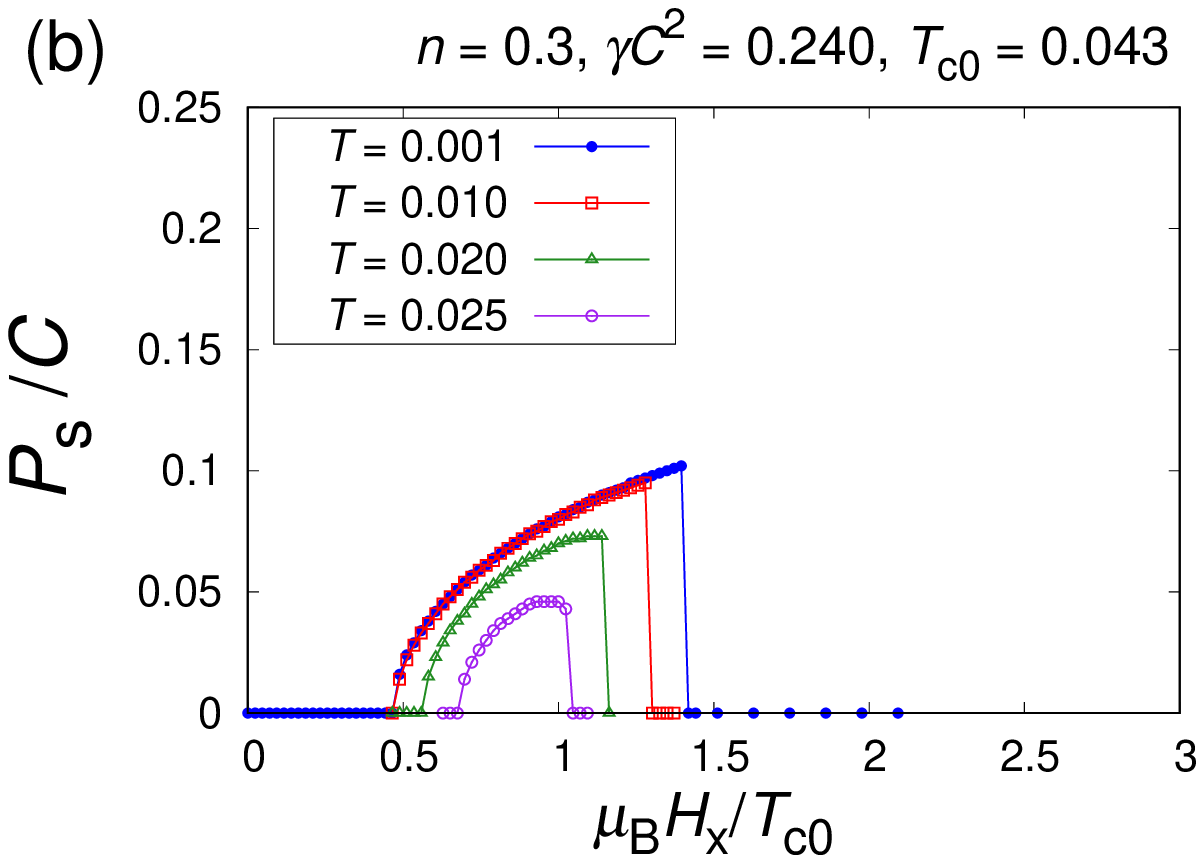}
  \end{minipage}
  \caption{\label{pol_metal}
 FE-like order parameter $P_s$ as a function of (a) the perpendicular field strength, and (b) the in-plane field strength for several values of $T$. 
}
\end{figure}

In order to discuss the order of the phase transition, we show the magnetic field dependence of the FE-like order parameter $P_s$ at several temperatures in Fig. \ref{pol_metal}. 
As shown in Fig. \ref{pol_metal}(a), the FE-like order parameter $P_s$ continuously increases with increasing the applied perpendicular field. 
This implies that the phase transition from the PE superconducting state to the FE superconducting state is second order. 
On the other hand, upon increasing the magnetic field after the FE-like structural transition occurs, an abrupt drop of $P_s$ occurs at the Pauli limiting field. 
Therefore, the transition from the FE superconducting state to the normal state or PE superconducting state by increasing the magnetic field is the first order phase transition. 
In Fig. \ref{phase_metal}, the type of phase transitions is illustrated by thickness of the transition lines. 

The responses to the magnetic field shown above are similar to the magnetoferroelectricity in multiferroic materials such as TbMnO$_3$ \cite{TbMnO3}, because the FE-like order is induced by the magnetic field in the superconducting state. 
Therefore, in the following, we call this magnetic-field-induced FE-like transition inside the superconducting state as {\it superconducting multiferroics}. 

%%%%%%%%%%%%%%%% FS %%%%%%%%%%%%%%%%%%%%%%%%%%%%%%%
%%%%%%%%%%%%%%%%%%%%%%%%%%%%%%%%%%%%%%%%%%%%%%%%%%%%%%%%%%%%

%%%%%%%%%%%%%%%% Pol %%%%%%%%%%%%%%%%%%%%%%%%%%%%%%%
%%%%%%%%%%%%%%%%%%%%%%%%%%%%%%%%%%%%%%%%%%%%%%%%%%%%%%%%%%%%

\subsection{\label{sec:level3C}In-plane magnetic field}
Here we turn into the in-plane magnetic field. 
Figure \ref{phase_metal}(b) shows the $H_x$-$T$ phase diagram in the high carrier density regime. 
The applied in-plane magnetic field stabilizes the FE superconducting state as we observed for the perpendicular field. 
On the other hand, unlike the case of the perpendicular field, the enhancement of the Pauli limiting field is small. 
This field angle dependence can also be understood as a result of the spin texture of Rashba superconductors. 
In-plane magnetic fields induce asymmetric deformation of the Fermi surfaces as shown in Fig. \ref{FS}(c). 
Although this deformation causes a suppression of the Pauli depairing effect, the Pauli limiting field at zero temperature is enhanced roughly by only a factor of $\sqrt{2}$, compared to centrosymmetric superconductors \cite{fujimoto2007electron}. 
Therefore, the FE superconducting state is stabilized under the in-plane field as well, and the Pauli limiting field is enhanced a little. 

Fig. \ref{pol_metal}(b) shows the the magnetic field dependence of $P_s$ at several values of $T$, and we can see that the order of the phase transition is the same as that for the perpendicular field. 
Since the suppression of the Pauli depairing effect is not significant in the in-plane field, the amplitude of $P_s$ is smaller than that for the perpendicular field. 

%%%% Sec. IV
\section{\label{sec:level4}Low carrier density regime}
\subsection{\label{sec:level4A}Dilute superconductivity}
Next, we study the superconducting phase diagram in the low carrier density regime by setting $n = 0.001$. 
This situation corresponds to the dilute superconductivity realized in doped STO. 
In such a low carrier density regime, qualitatively different behaviors from the high carrier density regime are shown below. 

\begin{figure}[t]
  \begin{minipage}[t]{0.45\linewidth}
    \centering
    \includegraphics[width=65mm,clip]{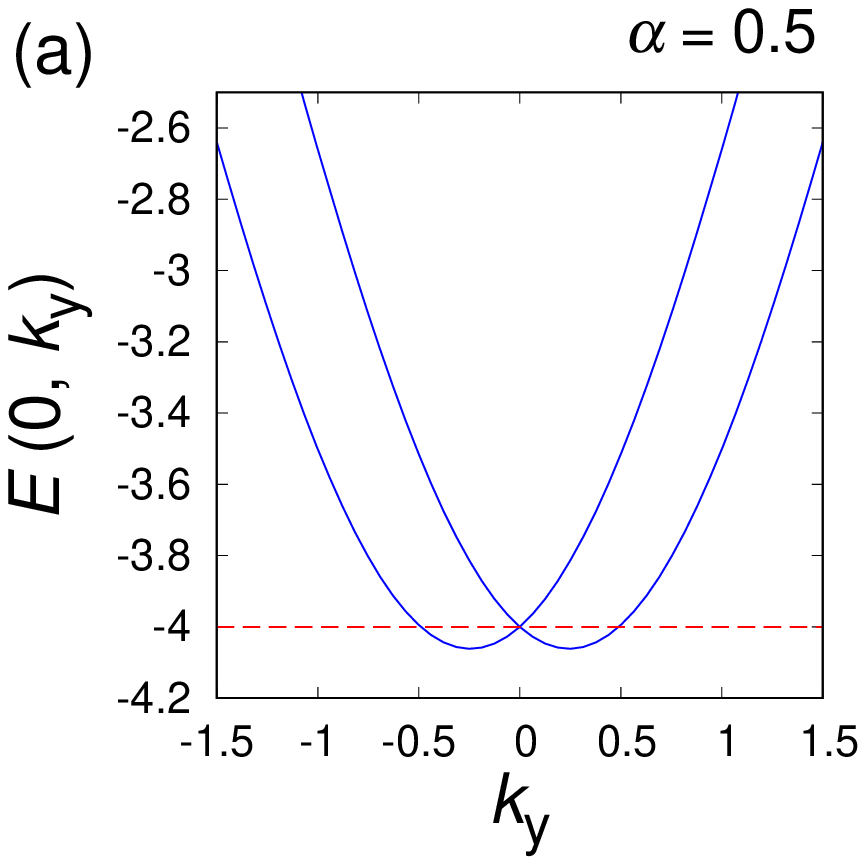}
  \end{minipage}
  \begin{minipage}[t]{0.45\linewidth}
    \centering
    \includegraphics[width=65mm,clip]{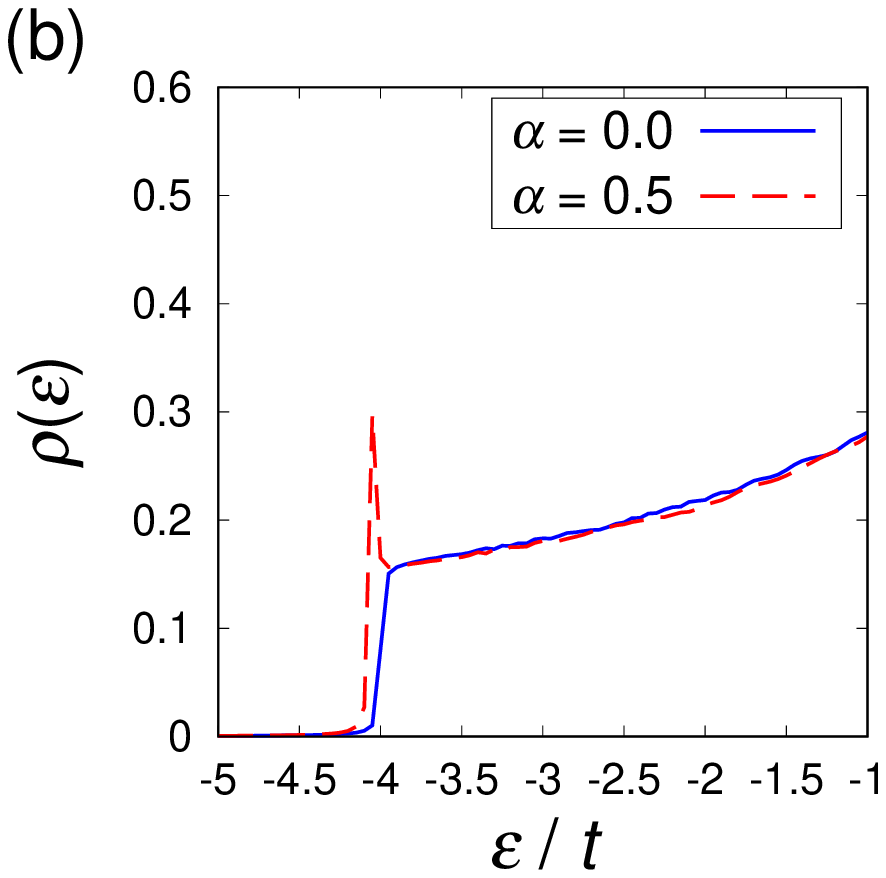}
  \end{minipage}
  \caption{\label{fig:low} 
(a) Band structure of a 2D Rashba system. The chemical potential is set to 
$\mu=0$ and a large ASOC $\alpha=0.5$ is assumed to emphasize the Rashba spin-splitting. The Lifshitz transition occurs at the dashed red line. 
(b) DOS $\rho(\varepsilon)$ in the normal state for $\alpha=0$ (solid line) and $\alpha=0.5$ (dashed line). 
The DOS is enhanced in the low carrier density regime for $\alpha=0.5$. 
The chemical potential is again set to $\mu=0$.  
}
\end{figure}

In order to elucidate properties of the dilute superconducting state, we first discuss the normal state in the low carrier density regime. 
Upon decreasing the carrier density for $\alpha \neq 0$, the Fermi energy becomes lower than the crossing point of the Rashba spin-split bands at the time-reversal invariant momentum $\vector{k}=\vector{0}$ [Fig. \ref{fig:low}(a)], and then, the nature of Rashba spin-split Fermi surfaces changes, {\it i.e.} the Lifshitz transition occurs. 
In a low carrier density regime below the critical value of the Lifshitz transition, the DOS significantly increases owing to an effective reduction of dimensionality \cite{PhysRevLett.98.167002}. 
On the other hand, when $\alpha=0$, the Lifshitz transition is accompanied by the metal-insulator transition, and the DOS vanishes after the Lifshitz transition. 
Therefore, the DOS is significantly increased by nonzero $\alpha$, namely, the ferroelectricity [Fig. \ref{fig:low}(b)]. 
This causes qualitatively different properties of the dilute superconductivity from high carrier density regime discussed in Sec. \ref{sec:level3}. 

We here consider the condensation energy $\Delta f = f_{\rm s} - f_{\rm n}$, which is the difference of free energy between the superconducting state and the normal state. 
According to the BCS theory, the condensation energy at zero temperature is obtained as follows:
\begin{equation}
\Delta f = f_{\rm s} - f_{\rm n} = - \frac{1}{2} \rho_0 |\Delta|^2, 
\label{eq:cond}
\end{equation}
where $\rho_0$ is the DOS per unit volume at the Fermi energy. 
The $\Delta$ non-linearly increases with $\rho_0$. 
Thus, the superconducting state is more stable for a larger $\rho_0$. 
Combining this to the above discussion on the normal state we expect that the FE superconducting state is more stable than the PE superconducting state in the low carrier density regime, because of the Rashba spin-splitting in the energy bands. 

%%%%%%%%%%%%%%%% low carrier %%%%%%%%%%%%%%%%%%%%%%%%%%%%%%%
%%%%%%%%%%%%%%%%%%%%%%%%%%%%%%%%%%%%%%%%%%%%%%%%%%%%%%%%%%%%

\subsection{\label{sec:level4B}Quantum ferroelectric criticality}
Here, we determine the value of $\gamma$ and $\eta$ in the same way as in Sec. \ref{sec:level3A}. 
In this section the magnetic field is set to zero. 
The attractive interaction is chosen to be $V_s/t=3.5$, and then the superconducting transition temperature $T_{\rm c0}$ is $T_{\rm c0}=0.018$ in the absence of the magnetic field. 
$\eta$ is chosen to be $\eta=0.003/C^4$ so that $\alpha < 0.5$. 
The following results are nearly independent of the value of $\eta$. 

\begin{figure}[t]
 \begin{center}
 \includegraphics[width=95mm]{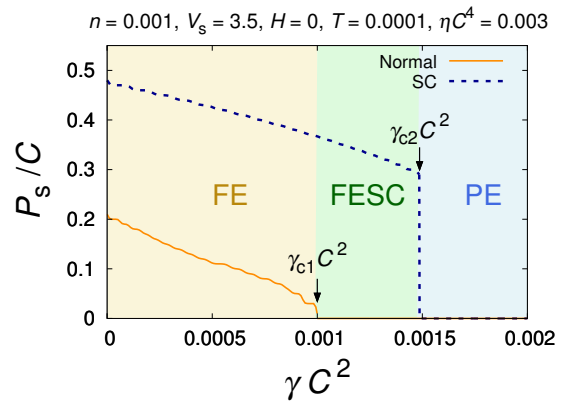}
 \end{center}
 \caption{\label{gamma_low} 
 $\gamma$ dependence of the FE-like order parameter 
$P_s$ in the low carrier density regime ($n=0.001$). 
The temperature is set to be $T=0.0001$, {\it i.e.} almost zero temperature. 
The blue dot line shows $P_s$ in the superconducting state, while the orange solid line shows $P_s$ in the normal state. 
Yellow and light blue area corresponds to the FE regime ($P_s \neq 0$) and PE regime ($P_s = 0$), respectively. 
Green area is the FE superconducting region in which the superconducting state is FE although the normal state is PE. 
}
\end{figure}

We show $\gamma$ dependence of $P_s$ in Fig. \ref{gamma_low}. 
We see that $P_s$ vanishes at $\gamma=\gamma_{\rm c1}=0.000998/ C^2$ when we assume non-superconducting state. 
On the other hand, the FE order survives until $\gamma$ exceeds 
$\gamma_{\rm c2}=0.001487/ C^2$ in the superconducting state. 
This result should be contrasted to the high carrier density regime where the two critical values $\gamma_{\rm c1}$ and $\gamma_{\rm c2}$ almost coincide. 
In the low carrier density regime, for a wide range of $\gamma$, the superconducting state is FE although the normal state is PE. 
In other words, the FE-like order occurs in the superconducting state without a polar lattice distortion in the normal state. 
As discussed in Sec. \ref{sec:level4A}, the condensation energy gained by the superconductivity is increased by the FE-like order in the low carrier density regime, and therefore, the FE superconducting state can be stabilized even in the absence of the magnetic field. 
This is a main result of this section. 
The existence of the wide FE superconducting region is an inherent nature of the dilute superconductivity, since it originates from 
the effective reduction of dimensionality in Rashba spin-split bands. 

%%%%%%%%%%%%%%%% QFE %%%%%%%%%%%%%%%%%%%%%%%%%%%%%%%
%%%%%%%%%%%%%%%%%%%%%%%%%%%%%%%%%%%%%%%%%%%%%%%%%%%%%%%%%%%%

%%%%%%%%%%%%%%%% Phase diagrams %%%%%%%%%%%%%%%%%%%%%%%%%%%%%%%
\begin{figure*}[t]
  \begin{minipage}[t]{0.45\linewidth}
    \centering
    \includegraphics[width=100mm,clip]{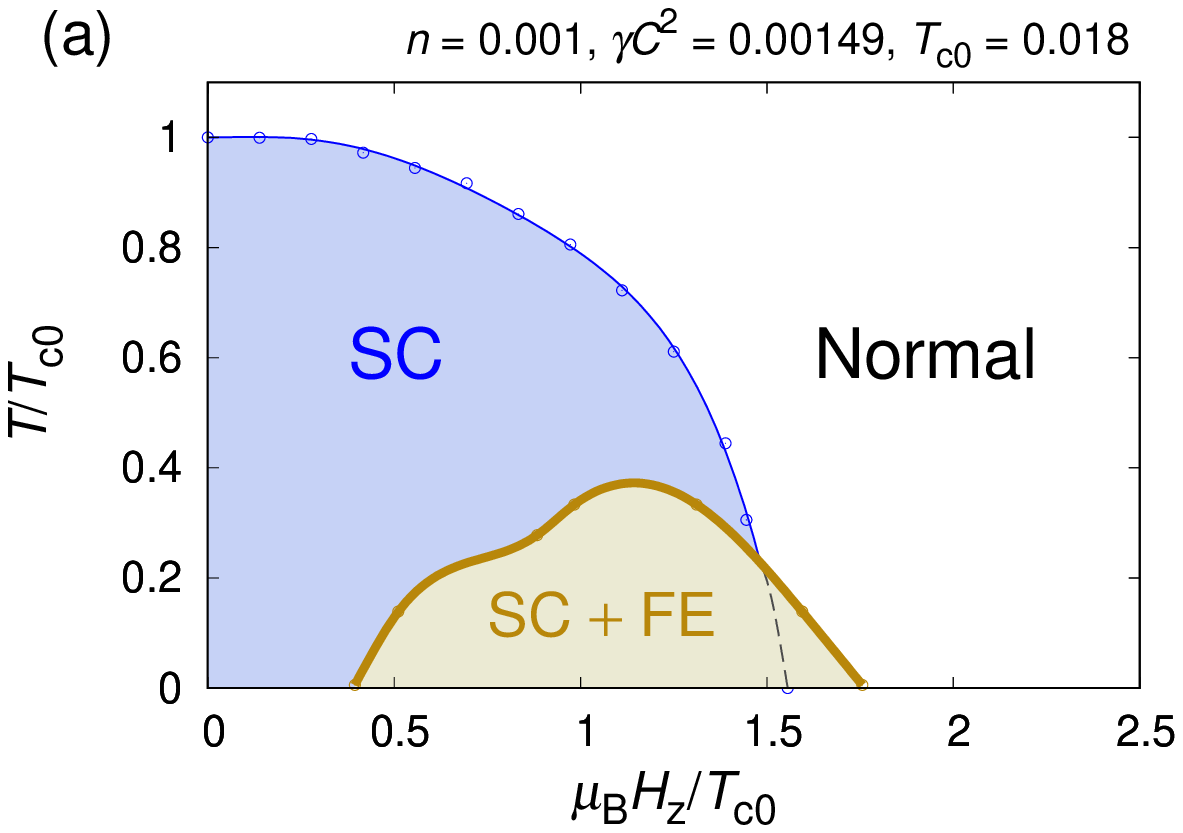}
  \end{minipage}
  \begin{minipage}[t]{0.45\linewidth}
    \centering
    \includegraphics[width=100mm,clip]{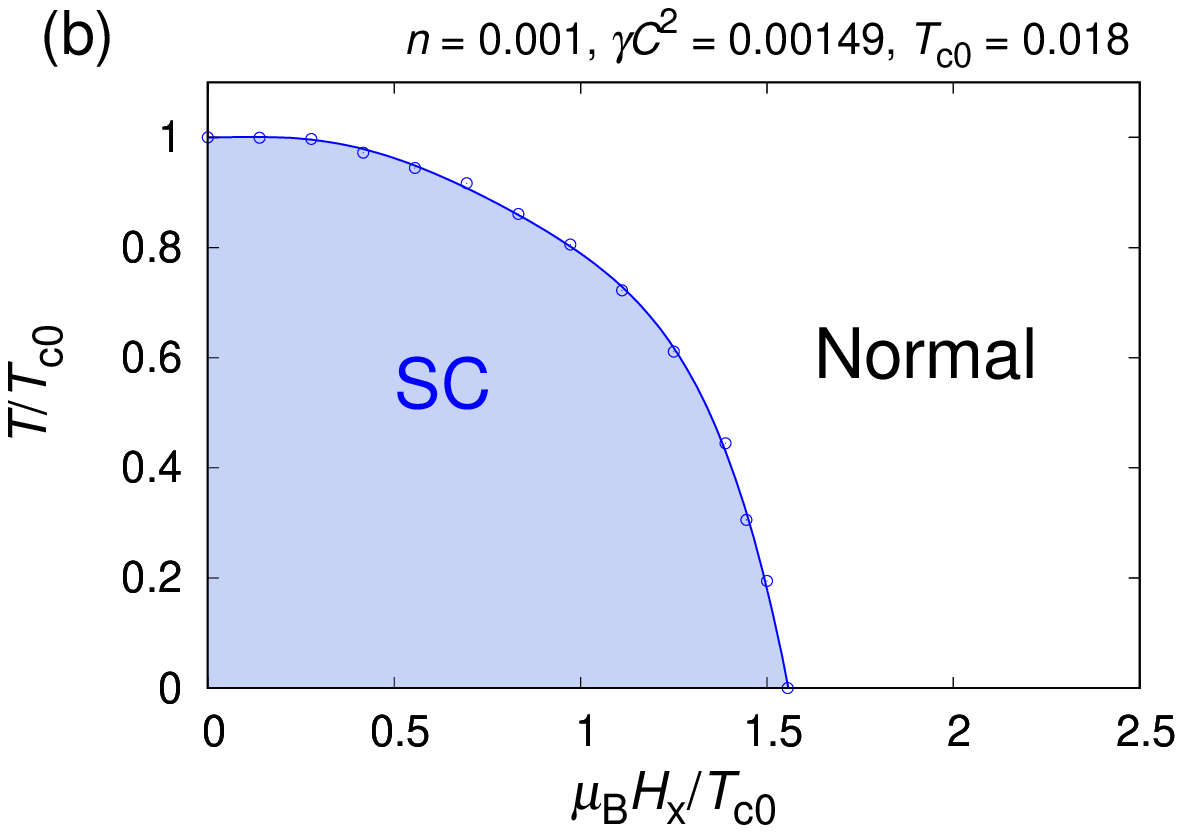}
  \end{minipage}
  \begin{minipage}[t]{0.45\linewidth}
    \centering
    \includegraphics[width=100mm,clip]{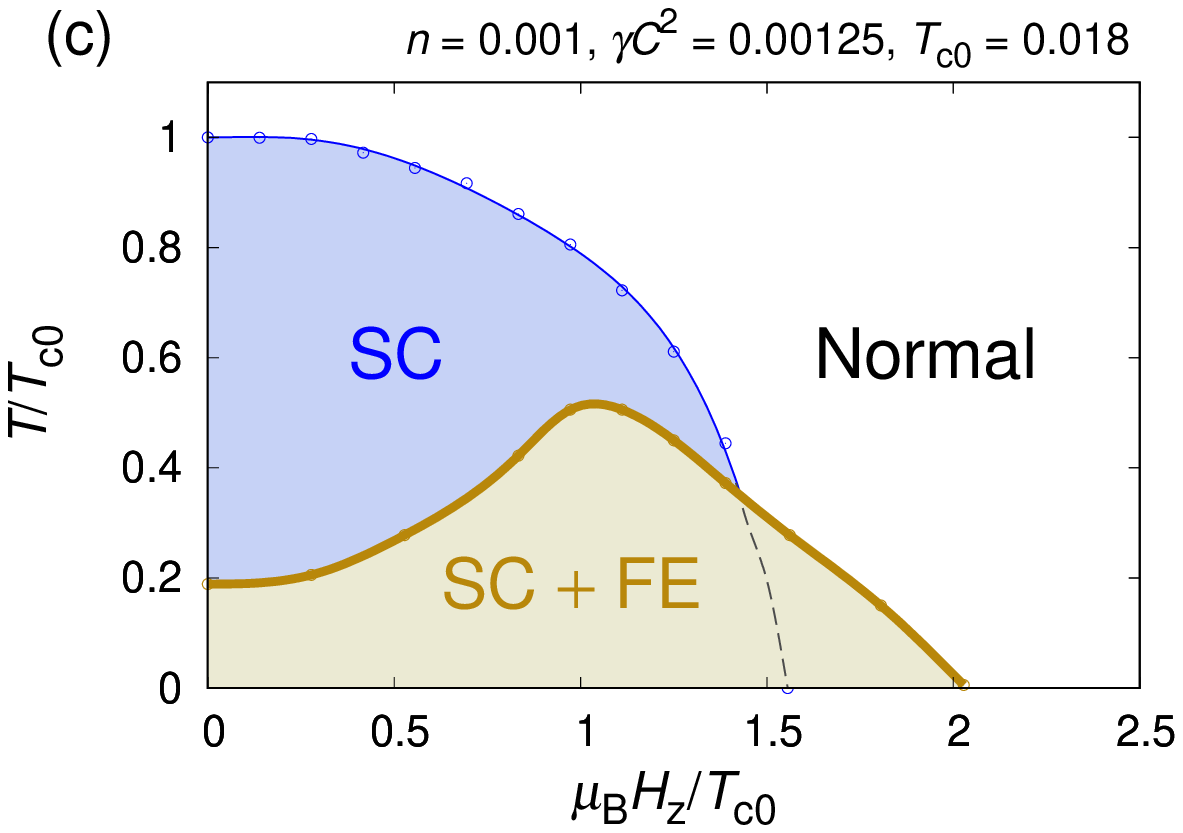}
  \end{minipage}
  \begin{minipage}[t]{0.45\linewidth}
    \centering
    \includegraphics[width=100mm,clip]{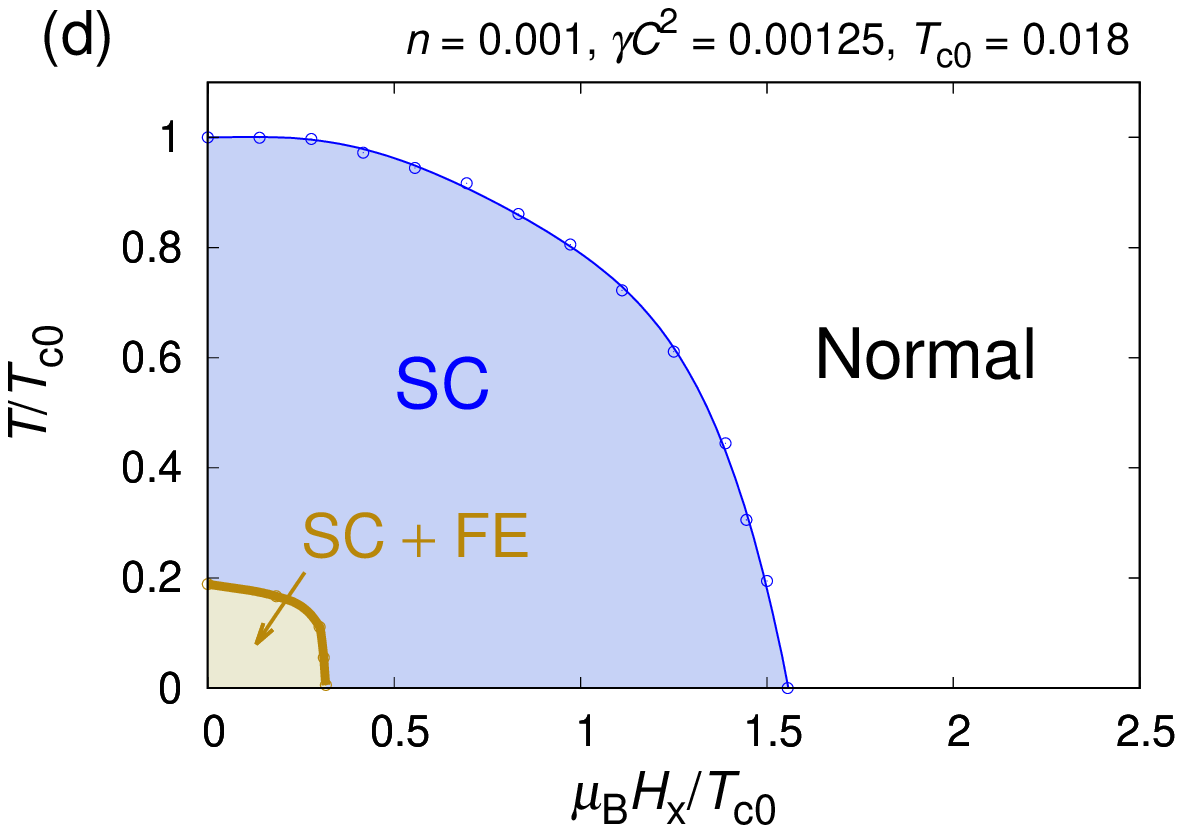}
  \end{minipage}
  \caption{\label{phase_low}
 Superconducting phase diagram in the low carrier density regime. 
(a) and (c) $T$-$H_z$ phase diagrams for (a) $\gamma=0.00149/ C^2$, and 
(c) $\gamma=0.00125/ C^2$. The black dashed line shows a superconducting phase transition line of a 2D superconductor with inversion symmetry. 
(b) and (d) $T$-$H_x$ phase diagrams for (b) $\gamma=0.00149 /C^2$, and 
(d) $\gamma=0.00125 /C^2$. 
The bold (thin) line shows the first-order (second-order) phase transition line. 
In contrast to the high carrier density regime, the FE-like structural transition is the first order phase transition in the whole phase diagram. 
The temperature $T$ and magnetic field $\mu_{\rm B}H$ are normalized by the superconducting transition temperature $T_{\rm c 0}$ at zero magnetic field. 
}
\end{figure*}
%%%%%%%%%%%%%%%%%%%%%%%%%%%%%%%%%%%%%%%%%%%%%%%%%%%%%%%%%%%%

\subsection{\label{sec:level4C}Perpendicular magnetic field}
In the following part we clarify effects of the magnetic field which are also qualitatively different from the high carrier density regime. 
We choose two values of $\gamma$. 
The first one is $\gamma=0.00149/ C^2$ slightly larger than the FE critical value $\gamma_{\rm c2}=0.001487 /C^2$. 
This situation is similar to Sec. \ref{sec:level3}. 
The other is $\gamma=0.00125 /C^2$ far below the critical point. 
In both cases the behaviors peculiar to the dilute superconductivity are shown below. 

Figures \ref{phase_low}(a) and \ref{phase_low}(c) show the $T$-$H_z$ phase diagram in the low carrier density regime for $\gamma=0.00149 /C^2$ 
and $\gamma=0.00125/ C^2$, respectively. 
For $\gamma=0.00149 /C^2$ ({\it i.e.}, slightly above the QCP at $H=0$), the FE superconducting state is stabilized under the perpendicular magnetic field, and the superconducting multiferroics is realized as it is in the high carrier density regime. 
On the other hand, for $\gamma=0.00125 /C^2$, the FE superconducting state is stabilized even at zero magnetic field. 
This result is consistent with the phase diagram of Sr$_{1-x}$Ca$_{x}$TiO$_{3-\delta}$ \cite{NatPhys.13.643-648}, in which the low carrier density regime is advantageous to host the FE superconducting phase, and the FE superconducting state is stabilized at zero magnetic field. 
Our results indicate that the FE-like order may occur in the superconducting state even in the doping region where the FE order disappears in the normal state. 
Figure \ref{phase_low}(c) also shows the enhancement of the FE superconductivity by magnetic field. 
The superconducting multiferroics in the low carrier density regime also arises from the suppression of the Pauli depairing effect in Rashba superconductors. 
Note that the change of the Fermi surfaces and spin texture induced by the perpendicular field is similar between the high carrier density regime and the low carrier density regime. 

\subsection{\label{sec:level4D}In-plane magnetic field}
Here we show that the response to the in-plane magnetic field is remarkably different between the low carrier density regime and the high density regime. 
Figures \ref{phase_low}(b) and \ref{phase_low}(d) show the $T$-$H_x$ phase diagram in the low carrier density regime for $\gamma=0.00149/ C^2$ and $\gamma=0.00125/ C^2$, respectively. 
For $\gamma=0.00149/ C^2$, the FE superconducting state is unstable in the whole phase diagram. 
Furthermore, for $\gamma=0.00125/ C^2$, the FE superconducting state is destabilized by applying the in-plane magnetic field. 
Thus, the behavior of the superconducting multiferroics is not observed in the low carrier density regime for the in-plane magnetic field. 

\begin{figure}[t]
  \begin{minipage}[t]{0.95\linewidth}
    \centering
    \includegraphics[width=80mm]{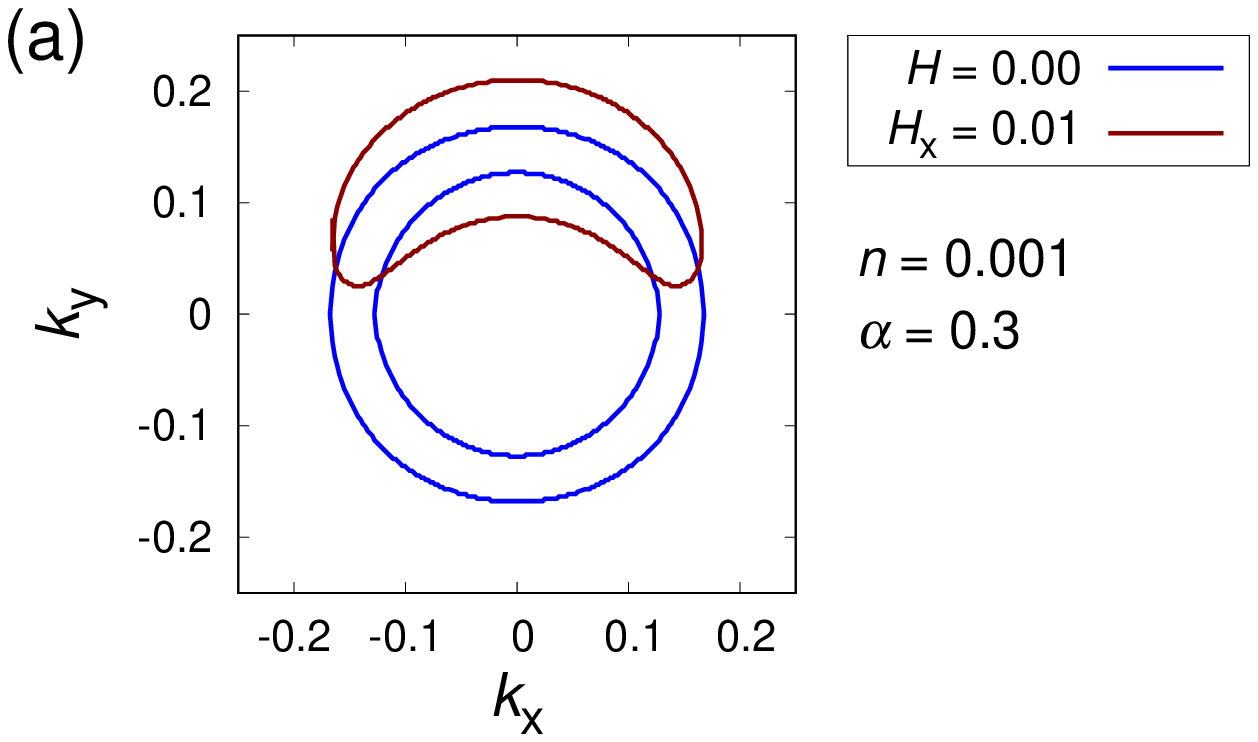}
  \end{minipage}
  \begin{minipage}[t]{1.0\linewidth}
    \centering
    \includegraphics[width=90mm]{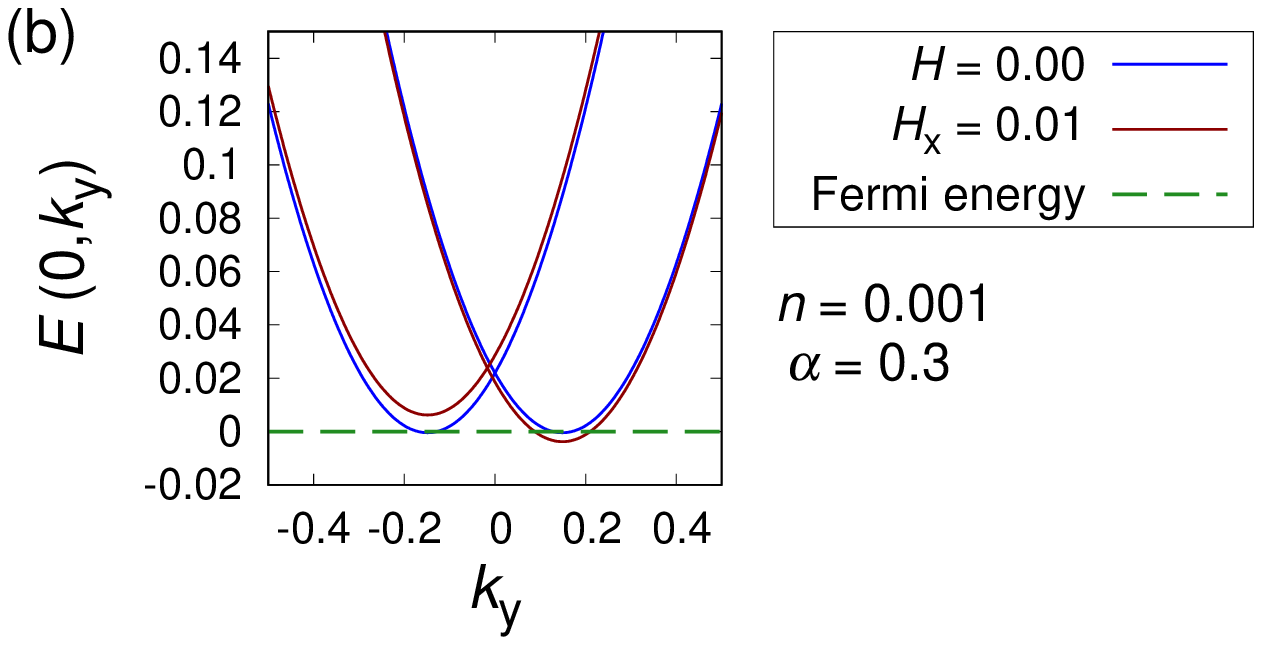}
  \end{minipage}
  \caption{\label{FS_low_hx}
 (a) Fermi surfaces in the normal state for 
$H_{x}=0$ (blue lines) and $H_{x}=0.01$ (red line). 
(b) The asymmetric deformation of Rashba spin-split bands under the in-plane field. The green dashed line illustrates the Fermi energy. 
The carrier density and Rashba spin-orbit coupling strength are set to $n=0.001$ and $\alpha=0.3$, respectively. 
}
\end{figure}

The difference from the high carrier density regime can be understood by noticing the drastic deformation of Fermi surfaces which is inherent in the low carrier density regime. 
As shown in Fig. \ref{FS_low_hx}(a), the topology of the Fermi surface drastically changes by applying the in-plane magnetic field, because of the asymmetric deformation of the band structure [Fig. \ref{FS_low_hx}(b)]. 
This Lifshitz transition and resulting asymmetric Fermi surface is destructive for the superconductivity. 
Highly asymmetric Fermi surface in Fig. \ref{FS_low_hx}(a) implies that spin-singlet Cooper pairs with $\vector{q}=\vector{0}$ assumed here cannot be generated. 
Therefore, the FE-like order competes with the superconductivity in the in-plane magnetic field while they are cooperative at $H=0$. 
Although the asymmetric band structure is ubiquitous in NCS systems under the magnetic field, such Lifshitz transition does not occur in the high carrier density regime. 
Therefore, the $T$-$H_x$ phase diagram in the low carrier density regime is qualitatively different from that in the high carrier density regime. 

On the other hand, the asymmetric Fermi surface makes the FFLO state with $\vector{q} \neq \vector{0}$ Cooper pairs more stable than the BCS state, and therefore, the destruction of the FE superconducting state might be prevented by stabilizing a FFLO state to some extent. 
This dilute FFLO state may be distinguished from the helical state which is realized in the high carrier density regime \cite{Dimitrova2003, PhysRevB.70.104521, PhysRevB.75.064511}, because of the Lifshitz transition of the Fermi surface in the low carrier density regime. 
Note that highly asymmetric Fermi surface in Fig. \ref{FS_low_hx}(a) also destabilize the FFLO state, and therefore, the phase diagram in Fig. \ref{phase_low}(d) is not qualitatively altered. 

%%%%%%%%%%%%%%%% FS_low_hx %%%%%%%%%%%%%%%%%%%%%%%%%%%%%%%
%%%%%%%%%%%%%%%%%%%%%%%%%%%%%%%%%%%%%%%%%%%%%%%%%%%%%%%%%%%%

\subsection{\label{sec:level4E}Topological superconductivity}
\begin{figure}[t]
 \begin{center}
 \includegraphics[width=95mm]{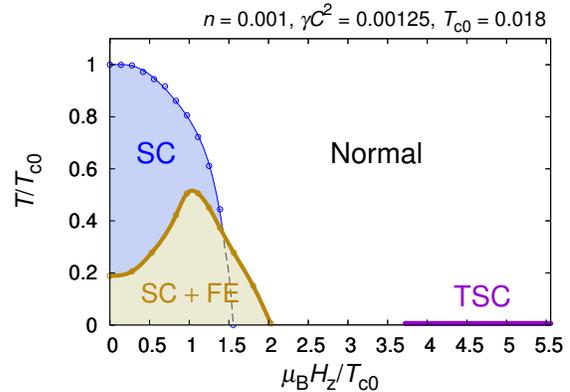}
 \end{center}
 \caption{\label{phase_topo} 
 $T$-$H_z$ phase diagram in the low carrier regime for 
$\gamma=0.00125 /C^2$ (the same figure as Fig. \ref{phase_low}(c)). 
Purple line shows the TSC region where the Chern number is nonzero if the FE superconducting state is stabilized. However, the superconductivity is unstable in this region. 
}
\end{figure}

Finally, we discuss a possibility of self-organized topological superconducting (TSC) phase 
\cite{PhysRevLett.111.186805, PhysRevLett.111.147202, PhysRevLett.111.206802, PhysRevB.93.140503, PhysRevB.93.134512} driven by the FE-like structural transition. 
It is known that the topologically nontrivial superconducting phase in D class, which is characterized by the nonzero Chern number, is realized in 2D Rashba superconductors by applying the perpendicular field \cite{PhysRevB.82.134521,PhysRevLett.103.020401}. 
Sato {\it et al.} \cite{PhysRevB.82.134521} clarified the condition of the topological superconductivity, and it has been shown that the low carrier density region is advantageous to realize the TSC phase. 
The FE superconducting state in our model satisfies the symmetry conditions, namely, broken inversion and time-reversal symmetry. 
Therefore, we may expect that the TSC state is self-organized in the low carrier density regime. 
However, we can show that the TSC state is thermodynamically unstable. 
Figure \ref{phase_topo} illustrates the TSC region in which the condition for nontrivial Chern number, 
$\sqrt{(4t+\mu)^2 + |\Delta|^2} < \mu_{\rm B} H_z < \sqrt{\mu^2 + |\Delta|^2}$, is satisfied. 
This region is outside the superconducting phase. 

The above result can be understood as follows. 
In order to realize the topological superconductivity in the dilute FE superconducting state, it is necessary to satisfy $\mu_{\rm B} H_z > |\Delta|$ at least. 
However, in such a weak superconducting region, the energy of lattice polarization is larger than the condensation energy gained by the superconductivity. 
Therefore, upon increasing the magnetic field, the FE superconducting state is destabilized before the topological phase transition occurs.

%%%%%%%%%%%%%%%% TOPO %%%%%%%%%%%%%%%%%%%%%%%%%%%%%%%
%%%%%%%%%%%%%%%%%%%%%%%%%%%%%%%%%%%%%%%%%%%%%%%%%%%%%%%%%%%%

%%%% Sec. V
\section{\label{sec:level5}Ginzburg-Landau theory}
In this section, we interpret our results based on the Ginzburg-Landau formulation, and provide a clear physical understanding of the spin-orbit-coupled FE superconductivity. 
By expanding Eq. (\ref{eq:free_BdG}) with respect to the two order parameters $\Delta$ and $P$ within quartic order, the electronic free energy $f_{\rm el}$ can be rewritten as follows:
\begin{eqnarray}
f_{\rm el} &\simeq& f_0 + a(T)|\Delta|^2 +\frac{1}{2} b(T) |\Delta|^4 \nonumber\\
&-& \frac{1}{2} \gamma_{\rm n}(T) P^2 
- \eta_{\rm n}(T) P^4 -
\frac{1}{2} \gamma_{\rm s}(T) |\Delta|^2  P^2 \nonumber\\ 
&+& \frac{1}{2}\left( \chi_{\rm n} - \chi_{\rm s}^{\perp} \right) H_z^2 
+ \frac{1}{2}\left( \chi_{\rm n} - \chi_{\rm s}^{\parallel} \right) H_x^2  . \label{eq:free_GL}
\end{eqnarray}
The first term $f_0$ is the free energy of PE normal state. 
According to the BCS theory, quadratic and quartic terms of the superconducting order parameter $\Delta$ are described as 
\begin{eqnarray}
a(T) &=& \rho_0 \left( \frac{T-T_{c0}}{T_{c0}}\right) , \\
b(T) &=& \rho_0 \frac{7 \zeta(3)}{8 (\pi T)^2} .
\end{eqnarray}
By differentiating Eq. (\ref{eq:free_BdG}) at $\Delta=0$ with respect to the $\alpha$, quadratic and quartic terms of the FE order parameter $P=C\alpha$ are obtained as
\begin{widetext}
\begin{eqnarray}
\gamma_{\rm n}(T) &=& \frac{\beta}{2 N C^2} \sum_{\bm{k}} |\vector{g}(\vector{k})|^2 \mathrm{sech}^2\left( \frac{\beta|\xi(\vector{k})|}{2} \right) , 
\label{eq:gamma_n} \\
\eta_{\rm n}(T) &=& \frac{\beta^3}{48N C^4} \sum_{\bm{k}} |\vector{g}(\vector{k})|^4 \mathrm{sech}^2\left( \frac{\beta|\xi(\vector{k})|}{2} \right) \left[ \tanh^2\left( \frac{\beta|\xi(\vector{k})|}{2} \right) - \frac{1}{2}
\mathrm{sech}^2\left( \frac{\beta|\xi(\vector{k})|}{2} \right)
\right] , \label{eq:eta_n}
\end{eqnarray}
\end{widetext}
where $\xi(\vector{k})=\varepsilon(\vector{k})-\mu$. 
Note that $\gamma_{\rm n}(T)$ is positive in the whole range of $T$. 
The coefficient of the coupling term between 
the superconducting and FE order parameters can be calculated in the same way, and obtained as 
\begin{widetext}
\begin{eqnarray}
\gamma_{\rm S}(T) =
\frac{1}{N C^2} \sum_{\bm k} 
\frac{|\bm{g}(\bm{k})|^2}{|\xi(\bm{k})|^3}
\left[ 
\tanh\left( \frac{\beta|\xi(\bm{k})|}{2} \right)
- \frac{\beta|\xi(\bm{k})|}{2} \mathrm{sech}^2\left( \frac{\beta|\xi(\bm{k})|}{2} \right)
\left\{
1+ \frac{\beta|\xi(\bm{k})|}{2}
\tanh\left( \frac{\beta|\xi(\bm{k})|}{2} \right)
\right\}
 \right] . \label{eq:gamma_s}
\end{eqnarray}
\end{widetext}
The last two terms of Eq. (\ref{eq:free_GL}) describe the magnetic energy due to the Zeeman coupling. 
$\chi_{\rm n} = 2 \mu_{\rm B} \rho_0$ is the magnetic susceptibility in the normal state. 
$\chi_{\rm s}^{\perp}$ and $\chi_{\rm s}^{\parallel}$ are the magnetic susceptibility in the superconducting state for the perpendicular and in-plane field, respectively. 
In Rashba superconductors, the magnetic susceptibility can be estimated as $\chi_{\rm s}^{\perp}=2\chi_{\rm s}^{\parallel} \sim \chi_{\rm n}(\alpha/T_{c0})^2(|\Delta|/T_{c0})^2$ \cite{fujimoto2007electron}, and hence the applied magnetic field induces the coupling term between 
superconducting and FE order parameters. 
Thus, the total free energy given by Eq. (\ref{eq:free_tot}) is rewritten as follows:
\begin{eqnarray}
f_{\rm tot} &\simeq& f_0 + a(T)|\Delta|^2 +\frac{1}{2} b(T) |\Delta|^4 \nonumber\\
&+& \frac{1}{2} \left(\gamma - \gamma_{\rm n}(T)\right) P^2 
+ \left( \eta - \eta_{\rm n}(T)\right) P^4  \nonumber\\ 
&-&  \frac{1}{2} \gamma_{\rm s}(T) |\Delta|^2  P^2
+ \frac{1}{2} \chi_{\rm n}\left( H_z^2+H_x^2 \right) \nonumber\\
&-& \frac{1}{2} \zeta \chi_{\rm n} \left( H_z^2 + \frac{1}{2} H_x^2 \right) \frac{P^2}{T_{c0}^2} 
\frac{|\Delta|^2}{T_{c0}^2}  , \label{eq:free_GL_tot}
\end{eqnarray}
where $\zeta>0$ is a constant. 

Using Eq. (\ref{eq:free_GL_tot}), we can get a clear understanding of the results  obtained in previous sections. 
First, we clarify the origin of the quantum FE criticality explained in Sec. \ref{sec:level3A} and \ref{sec:level4B}. 
Equation (\ref{eq:eta_n}) shows that $\eta_{\rm n}(T)$ is positive at almost zero temperature, and hence a non-zero $\eta$ is necessary to cutoff the value of $\alpha$ in a realistic regime. 
In addition, Eq. (\ref{eq:gamma_s}) shows that $\gamma_{\rm s}(T)$ is also positive at almost zero temperature. 
Therefore, the quantum FE transition occurs at $\gamma>0$ as shown in Figs. \ref{gamma_metal} and \ref{gamma_low}, and the critical values of $\gamma$ in the normal and superconducting state are estimated as $\gamma_{\rm c1}\simeq\gamma_{\rm n}$ and $\gamma_{\rm c2}\simeq\gamma_{\rm n}+\gamma_{\rm s} |\Delta|^2$, respectively. 
Since $\gamma_{\rm s}>0$, superconductivity stabilizes the FE order as shown in Figs. \ref{gamma_low} and \ref{phase_low}. 
Second, almost all of the magnetic responses of FE superconductor can also be understood from Eq. (\ref{eq:free_GL_tot}). 
Because of the last two terms of Eq. (\ref{eq:free_GL_tot}), the free energy of the superconducting state can be minimized at non-zero $P$ under the magnetic field even when it has a minimum at $P=0$ at zero magnetic field. 
On the other hand, the free energy in the normal state remains to be minimized at $P=0$ under the magnetic field. 
Thus, the magnetic field-driven FE superconductivity is realized. 
Furthermore, it is also obvious that the optimum value of $P_s$ for the in-plane field is smaller than that for the perpendicular field (see Fig. \ref{pol_metal}), since $\chi_{\rm s}^{\parallel} < \chi_{\rm s}^{\perp}$ is satisfied. 
However, the destabilization of the dilute FE superconducting state under the in-plane field [Figs. \ref{phase_low}(b) and \ref{phase_low}(d)] cannot be understood based on Eq. (\ref{eq:free_GL_tot}), since the effect of highly asymmetric deformation of the Fermi surface is not appropriately included in the Ginzburg-Landau formalism. 

%% %%Summary
\section{\label{sec:level6}Summary and discussion}
In this paper, we investigated the spontaneous coexistence of the superconductivity and the FE-like order, that is, the FE superconductivity. 
In a 2D tetragonal system with spin-orbit coupling, the Rashba-type ASOC arises as a result of a FE-like structural phase transition. 
Assuming linear relation between the polar lattice displacement and the coupling strength of the Rashba ASOC, we studied the 2D spin-orbit coupled system as a minimal model of FE superconductivity. 
The thermodynamically stable state was determined by calculating the free energy including the effect of the lattice polarization. 
The obtained results are summarized below. 

First, we have clarified the stability of the FE superconducting state in the high carrier density regime near the FE QCP. 
Because of the enhancement of the Pauli limiting field by Rashba ASOC, the FE superconducting state is stabilized by the applied magnetic field independent of the direction of the magnetic field. 
The FE superconducting region in the $T$-$H_z$ phase diagram is wider than that in the $T$-$H_x$ phase diagram, since the Pauli depairing effect for perpendicular field is suppressed more significantly than that for the in-plane field. 

Next, we have elucidated the superconductivity in the low carrier density regime. 
This situation may correspond to the dilute superconductivity in doped STO. 
As a result of the Lifshitz transition at the Dirac point of the Rashba spin-split bands, the FE superconducting state is stabilized even at zero magnetic field. 
This result is consistent with the recent experimental observation of FE superconductivity in Sr$_{1-x}$Ca$_{x}$TiO$_{3-\delta}$ \cite{NatPhys.13.643-648}. 
The magnetic field response is also different from that in the high carrier density regime. 
In contrast to the high carrier density case, the in-plane field destroys the FE-like order in the low carrier density regime, since the asymmetric Fermi surface under the in-plane field makes it difficult to generate the BCS pairing. 
This result implies the possibility of the dilute FFLO superconductivity which is different from the helical state in high carrier density Rashba superconductors. 
On the other hand, the perpendicular field enhances the stability of the FE superconducting state similarly to the high carrier density regime. 
Although the possibility of self-organized FE topological superconductivity has been examined, it has been shown that the TSC state is thermodynamically unstable in our model. 

Finally, we clarified the origin of our results based on the Ginzburg-Landau formulation. 
The total free energy of our model was expanded in respect to the superconducting and FE order parameter. 
The quantum FE criticality and the magnetic field-driven FE superconductivity were clearly understood by deriving the Ginzburg-Landau free energy. 
However, the $T$-$H_x$ phase diagram in the low carrier density regime could not be understood based on the Ginzburg-Landau formalism, because it is a consequence of the deformation of the Fermi surface which is not appropriately included in the Ginzburg-Landau free energy. 

This work is a proposal of magnetic-field-driven FE superconductivity in a spin-orbit-coupled system, namely, superconducting multiferroics, and we have revealed its ubiquitous mechanism by analyzing a minimal model. 
From our analysis, it is expected that the coexistent phase of superconductivity and ferroelectricity, which have been detected in Sr$_{1-x}$Ca$_{x}$TiO$_{3-\delta}$ \cite{NatPhys.13.643-648}, may be stabilized in various superconductors under the magnetic field. 
For instance, a megnetic-filed-induced polar lattice displacement may occur in superconducting $\delta$-doped STO \cite{Nature.462.487}, which is a 2D electron system described by our model. 
The superconducting multiferroic behavior is also expected in three-dimensional systems because the mechanism of FE superconductivity, namely, unusual magnetic response of NCS superconductors, is independent of the dimensionality \cite{fujimoto2007electron}. 
Therefore, SrTiO$_{3-\delta}$ may host the FE superconducting phase under the magnetic field, and the FE superconducting phase of Sr$_{1-x}$Ca$_{x}$TiO$_{3-\delta}$ might be broaden by the applied magnetic field. 
The quantum PE STO is a promising platform of the superconducting multiferroics. 
The zero-field FE superconductivity is also expected to occur in a dilute metallic state of STO. 
For more detailed studies of STO, an analysis of a three-dimensional and multi-orbital model is desired. 
Such a material-specific study of STO is left for a future work. 

Our theoretical proposal for the FE superconductivity is based on the unique magnetic field responses of NCS superconductors. 
Therefore, it is indicated that not only the FE-like structural transition, but also other types of inversion symmetry breaking order \cite{PhysRevLett.115.207002} might be induced in spin-orbit-coupled superconductors under applied magnetic field. 
For example, a pyrochlore oxide Cd$_{2}$Re$_2$O$_7$ \cite{hiroi2017pyrochlore, PhysRevB.95.020102} exhibits successive structural transitions under pressure, and becomes superconducting in the inversion symmetry broken phase \cite{PhysRevLett.87.187001, PhysRevB.64.180503, 0953-8984-13-33-105}. 
Another candidates are superconducting quasi-skutterudites Sr$_{3}T_{4}$Sn$_{13}$ ($T=$Rh, Ir) which undergo a structural transition from a cubic phase to a inversion broken phase \cite{PhysRevLett.109.237008, PhysRevLett.114.097002, PhysRevLett.115.207003, PhysRevB.95.155142}. 
Thus, an intriguing behavior, such as an enhancement of $T_{\rm c}$ or structural transition, can be induced by applying the magnetic field to these superconductors.

\begin{acknowledgments}
The authors are grateful to S. Sumita, H. Watanabe, J. Ishizuka and A. Daido for fruitful discussions. 
This work was supported by Grant-in-Aid for Scientific Research on Innovative Areas ``J-Physics'' (JP15H05884) and ``Topological Materials Science'' (JP16H00991) from JSPS of Japan, and by JSPS KAKENHI Grants No. JP15K05164 and No. JP15H05745. 
\end{acknowledgments}

\nocite{*}

\providecommand{\noopsort}[1]{}\providecommand{\singleletter}[1]{#1}%

\end{document}